\documentstyle[12pt]{article}
\global\arraycolsep=2pt 

\newcommand{\leqsim}{\,\mbox{{\scriptsize $\stackrel{<}{\sim}$}}\,}

\newcommand{\beq}{\begin{equation}}
\newcommand{\eeq}{\end{equation}}
\newcommand{\bea}{\begin{eqnarray}}
\newcommand{\eea}{\end{eqnarray}}

\input{epsf}

\begin{document}

\thispagestyle{empty}

\rightline{TTP97-22}
\rightline{hep-ph/9706261}
\rightline{June 1997}
\bigskip
\boldmath
\begin{center}
{\Large{\bf New physics in penguin dominated $B\to\pi K$ decays}}
\vspace*{0.3truecm}
\end{center}
\unboldmath
\smallskip
\begin{center}
{\large{\sc Robert Fleischer}}\footnote{Internet: {\tt 
rf@ttpux1.physik.uni-karlsruhe.de}}\, and \, {\large{\sc Thomas 
Mannel}}\footnote{Internet: {\tt 
tm@ttpux7.physik.uni-karlsruhe.de}}\\
\vspace{0.3cm}
{\sl Institut f\"{u}r Theoretische Teilchenphysik}\\
{\sl Universit\"{a}t Karlsruhe}\\
{\sl D--76128 Karlsruhe, Germany}\\ 
\vspace{0.3cm}
\end{center}
\begin{abstract}
\noindent
Measurements of the combined branching ratios for $B^\pm\to\pi^\pm K$ and 
$B_d\to\pi^\mp K^\pm$ allow interesting constraints on the CKM angle $\gamma$,
the ratio $r\equiv|T'|/|P'|$ of the current-current and penguin 
operator contributions to $B_d\to\pi^\mp K^\pm$, and the CP asymmetry  
in that decay. Present CLEO results for these branching ratios indicate
where problems with consistency of the Standard Model may arise in 
the future. In this paper we discuss scenarios of new physics 
in these decays and investigate their implications for the above 
constraints.
\end{abstract}

\section{Introduction}
Using flavor symmetries of strong interactions, decays of $B$ mesons into 
$\pi K$ and $\pi\pi$ final states play an important role to determine the 
angles of the unitarity triangle \cite{ut} of the CKM matrix \cite{ckm}, 
in particular for the angle $\gamma$ which is notoriously difficult to 
measure at $B$ factories (see e.g.\ \cite{rev} for a recent review). 
An experimentally promising approach to determine $\gamma$ with 
the help of the branching ratios for $B^+\to\pi^+K^0$, $B^0_d\to\pi^-K^+$ 
and their charge-conjugates was proposed in \cite{PAPIII}. Recently the CLEO 
collaboration has reported a first measurement of these decays \cite{CLEO}. 
Since at present only results for the combined branching ratios 
\begin{eqnarray}
\mbox{BR}(B^\pm\to\pi^\pm K)&\equiv&\frac{1}{2}\left[\mbox{BR}(B^+\to\pi^+K^0)
+\mbox{BR}(B^-\to\pi^-\overline{K^0})\right]\label{BR-char}\\
\mbox{BR}(B_d\to\pi^\mp K^\pm)&\equiv&\frac{1}{2}
\left[\mbox{BR}(B^0_d\to\pi^-K^+)
+\mbox{BR}(\overline{B^0_d}\to\pi^+K^-)\right]\label{BR-neut}
\end{eqnarray}
are available, it is unfortunately not yet possible to fix $\gamma$ using 
that approach. 

However, as we have pointed out in a recent paper \cite{fm2}, even the 
{\it combined} branching ratios (\ref{BR-char}) and (\ref{BR-neut}) allow to 
derive stringent constraints on $\gamma$. So far information about that
angle could only be obtained in an indirect way by using experimental data 
on $|V_{cb}|$, $|V_{ub}|/|V_{cb}|$, $B^0_d$--$\overline{B^0_d}$ mixing and 
CP violation in the neutral $K$-meson system. Following these lines and using
present data, one typically finds in the SM framework 
(see e.g.\ \cite{bf-rev,al})
\begin{equation}\label{gamma-normal}
40^\circ \leqsim \gamma \leqsim 140^\circ.
\end{equation}

Using on the other hand our approach \cite{fm2}, one gets an allowed
range for $\gamma$ that is complementary to (\ref{gamma-normal}) and is given 
by
\begin{equation}\label{gamma-bound1}
0^\circ\leq\gamma\leq\gamma_0\quad\lor\quad180^\circ-
\gamma_0\leq\gamma\leq180^\circ\,,
\end{equation}
where $\gamma_0$ is related both to 
\begin{equation}\label{R-det}
R=\frac{\mbox{BR}(B_d\to\pi^\mp K^\pm)}{\mbox{BR}(B^\pm\to
\pi^\pm K)}
\end{equation}
and to the amplitude ratio 
\begin{equation}\label{Def-r}
r\equiv\frac{|T'|}{|P'|}
\end{equation}
of the current-current and penguin operator contributions to $B_d\to
\pi^\mp K^\pm$. The consistent description of $B^\pm\to\pi^\pm K$ and 
$B_d\to\pi^\mp K^\pm$ within the Standard Model (SM) implies furthermore
the allowed range 
\begin{equation}\label{range-r}
\left|1-\sqrt{R}\right|\leq r\leq1+\sqrt{R}
\end{equation}
and upper limits for the CP asymmetry arising in $B_d\to\pi^\mp K^\pm$.
It is interesting to note that commonly accepted means to estimate $r$ 
yield rather small values that are at the edge of compatibility with the 
present CLEO results due to the lower bound in (\ref{range-r}). 

Concerning the constraints (\ref{gamma-bound1}) on $\gamma$, an important 
difference arises between $R<1$ and $R>1$. In the former case -- the central 
values of the present CLEO results \cite{CLEO}
\begin{eqnarray}
\mbox{BR}(B^\pm\to\pi^\pm K)&=&\left(2.3^{+1.1+0.2}_{-1.0-0.2}
\pm0.2\right)\cdot
10^{-5}\label{BR-charres}\\
\mbox{BR}(B_d\to\pi^\mp K^\pm)&=&\left(1.5^{+0.5+0.1}_{-0.4-0.1}
\pm0.1\right)\cdot
10^{-5}\label{BR-neutres}
\end{eqnarray}
give $R=0.65<1$ -- the bound $\gamma_0$ takes a maximal value
\begin{equation}
\gamma_0^{\rm max} = \arccos \left( \sqrt{1-R} \right)
\end{equation}
independent of the amplitude ratio $r$. To be specific, for $R=0.65$ we 
have $\gamma_0^{\rm max}=54^\circ$. The CKM angle $\gamma$ can even be 
constrained in a more restrictive way if one uses additional knowledge on
$r$. In contrast to $R<1$, if $R$ is found to be larger than 1, such 
information on $r$ is required to constrain $\gamma$. 

Consequently, once more data come in confirming $R<1$, the SM can be put to 
a decisive test and our approach \cite{fm2} could give hints for ``New
Physics''. Therefore it is an interesting issue to analyze new physics effects 
in $B\to\pi K$ decays and in particular their implications for the bounds 
derived in \cite{fm2}. Such considerations are the topic of the present 
paper. In Section~\ref{param} we parametrize contributions of physics beyond 
the SM to the modes $B^\pm\to\pi^\pm K$ and $B_d\to\pi^\mp K^\pm$. Using
some specific scenarios of new physics, we investigate the corresponding 
modifications of the constraints arising from these decays in 
Section~\ref{disc}. The main results are summarized briefly in 
Section~\ref{concl}.

\boldmath
\section{Parametrization of new physics in $B\to\pi K$}\label{param}
\unboldmath
Before we turn to effects of new physics, let us recall the general structure 
of the SM transition amplitudes for $B^\pm\to\pi^\pm K$ and $B_d\to\pi^\mp 
K^\pm$. Using $SU(2)$ isospin symmetry of strong interactions to relate the 
hadronic matrix elements of the relevant four-quark operators, we may write 
these amplitudes as \cite{PAPIII,fm2}
\begin{eqnarray}
A(B^+ \to \pi^+K^0) &=& P'-\frac{1}{3} P_{{\rm EW}}^{\prime{\rm C}} 
\label{Ampl-char}  \\
A(B^0_d \to\pi^- K^+) &=& - \left[\left(P'+\frac{2}{3}P_{{\rm 
EW}}^{\prime{\rm C}}\right) + T'\right], \label{Ampl-neut}
\end{eqnarray}
where $P'$ and $P_{{\rm EW}}^{\prime{\rm C}}$ denote the QCD and 
color-suppressed electroweak (EW) penguin amplitudes, respectively, and 
$T'$ is the color-allowed $\bar b\to\bar uu\bar s$ current-current 
amplitude. Estimates within the SM yield \cite{rev,fm2} 
\begin{equation}\label{rho-EWP}
\rho_{\rm EWP}\equiv\frac{|P_{{\rm EW}}^{\prime{\rm C}}|}{|P'|}={\cal 
O}(10^{-2})\,,
\end{equation}
so that the EW penguin amplitudes in (\ref{Ampl-char}) and 
(\ref{Ampl-neut}) are expected to play a very minor role \cite{rev}. 
Consequently we will
neglect these contributions in the following discussion as we have done 
in \cite{fm2}. In Subsection~\ref{EWPs} we shall come back to this issue, 
assuming a large enhancement of the EW penguins.

Since we are considering only $B^\pm\to\pi^\pm K$ and $B_d\to\pi^\mp 
K^\pm$ decays, new physics can be incorporated very generally 
by modifying the two amplitudes 
(\ref{Ampl-char}) and (\ref{Ampl-neut}). To this end we write 
in a completely general way
\begin{eqnarray}
A(B^+ \to \pi^+K^0) &=& P' + P^{\rm new}_d
\label{new-char}  \\
A(B^0_d \to\pi^- K^+) &=& - \left(P' + T' + P^{\rm new}_u\right), 
\label{new-neut}
\end{eqnarray}
where $P^{\rm new}_d$ and $P^{\rm new}_u$ are the new physics contributions
to $\bar{b} \to \bar{s} d \bar{d}$ and $\bar{b} \to \bar{s} u \bar{u}$
quark-level transitions, respectively. 

We are obviously not in a position to fix the two complex amplitudes
from present data. This means that additional assumptions have to be made in 
order to reduce the number of unknown parameters. In principle one 
could refer now to commonly used models for physics beyond the SM
to estimate the new physics contributions \cite{NewPhysModels,appls}. Here 
we shall simply use some generic assumptions related to the behaviour 
of the new physics under isospin. Although these additional assumptions 
may appear ad hoc, we think that they are of at least comparable use as 
the model estimates performed before. 
 
Our first assumption is that no direct CP violation shows up in the 
new physics contributions to the decays under consideration, i.e.\
\begin{equation}\label{no-CP}
|\overline{P^{\rm \,new}_q}|=|P^{\rm new}_q|\,,
\end{equation}
where $q\in\{d,u\}$ and the overlined amplitudes correspond to the 
charge-conjugate processes. In that particular case the $P^{\rm new}_q$
amplitudes can be expressed as
\begin{equation}\label{pen-ampl}
P^{\rm new}_q=e^{i\phi_q}e^{i\delta^{\rm new}_q}|P^{\rm new}_q|\,,
\end{equation}
where $\delta^{\rm new}_q$ is a CP-conserving strong phase and 
$\phi_q$ a CP-violating weak phase so that we have
\begin{equation}
\overline{P^{\rm \,new}_q}=e^{-i\phi_q}e^{i\delta^{\rm new}_q}
|P^{\rm new}_q|\,.
\end{equation}
Taking into account (see e.g.\ \cite{rev,fm2}) 
\begin{equation}\label{SM-amplitudes}
P'=-e^{i\delta_{P'}}|P'|=\overline{P'},\quad 
T'=e^{i\gamma}e^{i\delta_{T'}}|T'|,\quad
\overline{T'}=e^{-i\gamma}e^{i\delta_{T'}}|T'|
\end{equation}
and using (\ref{pen-ampl}), we get 
\begin{eqnarray}
\left\langle|A(B^\pm\to\pi^\pm K)|^2\right\rangle&=&|P'|^2\left(1-
2\,\rho_d\cos\Delta_d\cos\phi_d+\rho_d^2\right)\\
\left\langle|A(B_d\to\pi^\mp K^\pm)|^2\right\rangle&=&|P'|^2\Bigl[1-2\,r
\cos\delta\cos\gamma-2\,\rho_u\cos\Delta_u\cos\phi_u\nonumber\\
&& +2\,\rho_u\,r\,\cos(\Delta_u-
\delta)\cos(\phi_u-\gamma)+\rho_u^2+r^2\Bigr]\,,
\end{eqnarray}
where the ``averages'' are defined by $\langle|A|^2\rangle\equiv(|A|^2+
|\overline{A}|^2)/2$, $\delta\equiv\delta_{T'}-\delta_{P'}$ and $\Delta_q
\equiv\delta^{\rm new}_q-\delta_{P'}$ denote differences of CP-conserving 
strong phases, and the parameters $\rho_q\equiv|P^{\rm new}_q|/|P'|$ measure 
the strengths of the new physics contributions relative to the QCD penguin 
amplitude. 

We do not expect the quantities $\rho_q$ to be small of the order $M_W^2 / 
\Lambda^2$, where $\Lambda^2$ is the scale of new physics. While such 
a suppression is active for CKM allowed tree level processes,  in our 
case $\rho_q={\cal O}(0.5)$ is not unreasonable due to the loop 
suppression of the QCD penguins in the SM.

A striking effect of new physics in $B^\pm\to\pi^\pm K$ would be a large 
CP asymmetry
\begin{equation}\label{CP-asym1}
{\cal A}_{\rm CP}^{\rm dir}(B^+\to\pi^+K^0)\equiv\frac{\mbox{BR}
(B^+\to\pi^+K^0)-\mbox{BR}(B^-\to\pi^-\overline{K^0})}{\mbox{BR}
(B^+\to\pi^+K^0)+\mbox{BR}(B^-\to\pi^-\overline{K^0})}\,.
\end{equation}
Within the SM only very small values of that asymmetry, at most of 
${\cal O}(1\%)$ \cite{pens}, can be accommodated\footnote{These tiny 
effects are neglected in our formulae.}, whereas interference 
between the QCD penguin and new physics contributions may lead to potentially 
large CP-violating effects in that decay which are described by
\begin{equation}\label{CP-asym1-new}
{\cal A}_{\rm CP}^{\rm dir}(B^+\to\pi^+K^0)=
\frac{2\,\rho_d\sin\Delta_d\,\sin\phi_d}{1-2\,\rho_d\cos\Delta_d\,\cos\phi_d+
\rho_d^2}
\end{equation}
and require that both $\Delta_d$ and $\phi_d$ take values different from 0
or $\pi$.

One of the central ingredients of our approach \cite{fm2} to constrain 
$\gamma$ is the quantity
\begin{equation}\label{DEF-R}
R\equiv\frac{\left\langle|A(B_d\to\pi^\mp K^\pm)|^2\right\rangle}{|P'|^2}\,,
\end{equation}
which is given within the SM, neglecting small phase-space and $B$ lifetime
differences, by the ratio (\ref{R-det}). That is, however,
not the case in the presence of new physics. In order to distinguish 
(\ref{R-det}) from (\ref{DEF-R}), we refer to the former ratio in the 
following discussion 
as $R_{\rm exp}$ since it can be obtained directly from the combined 
branching ratios that have been specified in (\ref{BR-char}) and 
(\ref{BR-neut}). Another important quantity is the amplitude ratio 
(\ref{Def-r}). In \cite{fm2} we were using the combined branching ratio for 
$B^\pm\to\pi^\pm K$ to fix the magnitude of the QCD penguin amplitude $P'$ 
yielding
\begin{equation}\label{DEF-rexp}
r=\frac{|T'|}{\sqrt{\left\langle|A(B^\pm\to\pi^\pm K)|^2\right\rangle}}\,.
\end{equation}
In the presence of new physics, the right-hand side of that equation does not 
measure (\ref{Def-r}). Since that ratio will nevertheless play an important 
role for our considerations, we refer to (\ref{DEF-rexp}) in the following as 
$r_{\rm exp}$. Strategies to fix $|T'|$ are discussed in \cite{fm2}.

In order to derive constraints from the combined $B\to\pi K$ branching ratios
(\ref{BR-char}) and (\ref{BR-neut}), the combination
\begin{equation}\label{C-def}
C\equiv\frac{1-R}{2\,r}+\frac{1}{2}\,r
\end{equation}
of $R$ and $r$ plays a central role as we have pointed out in \cite{fm2}. 
Within the SM, this quantity is simply given by the product of $\cos\delta$ 
and $\cos\gamma$. Taking into account possible contributions from new physics, 
we get
\begin{equation}\label{Def-cnew}
C_{\rm new}\equiv\frac{1-(R_{\exp}+R_{\rm new})}{2\,r_{\rm exp}}+\frac{1}{2}
\,r_{\rm exp}=\frac{\cos\delta\cos\gamma-\rho_u\cos(\Delta_u-\delta)
\cos(\phi_u-\gamma)}{\sqrt{1-2\,\rho_d\cos\Delta_d\cos\phi_d+\rho_d^2}}\,,
\end{equation}
where
\begin{equation}\label{Def-rnew}
R_{\rm new}\equiv\frac{2\,(\rho_u\cos\Delta_u\cos\phi_u-\rho_d\cos\Delta_d
\cos\phi_d)+\rho_d^2-\rho_u^2}{1-2\,\rho_d\cos\Delta_d\cos\phi_d+\rho_d^2}\,.
\end{equation}
Consequently new physics manifests itself in two ways: first the relevant 
value of $R$ is shifted from its measured value $R_{\exp}$ by $R_{\rm new}$, 
and second $C_{\rm new}$ is no longer related 
in a simple way to $\cos\gamma$, i.e.\ to the weak CP-violating phase of 
the $T'$ amplitude.  
Let us note that our expressions are still very general since we have 
so far only used (\ref{no-CP}) to simplify our analysis. 

\boldmath
\section{Scenarios of new physics in $B\to\pi K$ decays}\label{disc}
\unboldmath
In order to proceed further, we have to make additional assumptions to 
reduce the number of unknown parameters. To this end we will focus on 
some scenarios of new physics. A very transparent one is discussed in the 
following subsection.

\boldmath
\subsection{New physics I: $SU(2)$ isospin-symmetric case}
\unboldmath
One of the basic assumptions in this subsection is that the new physics 
contributions are equal for $B^\pm\to\pi^\pm K$ and $B_d\to\pi^\mp K^\pm$,
i.e.\ couple equally to $d$- and $u$-quarks, so that we have
\begin{equation}\label{isospin}
\begin{array}{lcrcr}
\phi_d&=&\phi_u&=&\phi\\
\rho_d&=&\rho_u&=&\rho\\
\Delta_d&=&\Delta_u&=&\Delta
\end{array}
\end{equation}
implying $R_{\rm new}=0$. In fact, isospin is conserved in many new 
physics scenarios, such as models with enhanced chromomagnetic dipole 
operators \cite{kagan}.

In addition we assume that the strong phase 
difference between the QCD penguin and the new physics contributions 
vanishes, i.e.\ $\Delta=0$. Consequently the CP asymmetry 
(\ref{CP-asym1-new}) for $B^+\to\pi^+K^0$ is zero in that case as in the 
SM. Combining all these assumptions we get
\begin{equation}\label{C-iso}
C_{\rm new}=\frac{1-R_{\exp}}{2\,r_{\rm exp}}+\frac{1}{2}
\,r_{\rm exp}=\cos\delta\,\cos\gamma_{\rm exp}
\end{equation}
and
\begin{equation}\label{CPasym}
{\cal A}_{\rm CP}^{\rm dir}(B_d^0\to\pi^-K^+)=2\,\frac{r_{\rm exp}}
{R_{\rm exp}}\,\sin\delta\,\sin\gamma_{\rm exp}\,,
\end{equation}
where
\begin{equation}
\gamma_{\rm exp}=\gamma+\Gamma
\end{equation}
with
\begin{equation}
\cos\Gamma=\frac{1-\rho\,\cos\phi}{\sqrt{1-2\,\rho\,\cos\phi+
\rho^2}}\,,\quad
\sin\Gamma=\frac{\rho\,\sin\phi}{\sqrt{1-2\,\rho\,\cos\phi+\rho^2}}\,.
\end{equation}
Therefore the experimentally determined angle $\gamma_{\rm exp}$ is
not equal to the weak phase $\gamma$ of the $T'$ amplitude, but is shifted 
by $\Gamma$. 

There are several strategies for experimental determinations 
of $\gamma$ on the market \cite{rev}.
Within the SM, these methods would all yield the same 
value of $\gamma$. Once new physics shows up, differences may appear between 
these results since one type of strategies refers to charged $B$ decays 
originating from $b \to c\bar us$ ($\bar b\to\bar uc\bar s$) transitions 
that receive only current-current and no penguin contributions~\cite{gw}, 
while another type uses $B_s^0-\overline{B_s^0}$ mixing \cite{hawaii}. 
Since new physics is expected to affect these mixing processes significantly 
-- in particular the corresponding mixing phase -- as they are 
suppressed FCNC loop processes \cite{NewPhysModels}, the latter methods 
are sensitive to physics beyond the SM similarly as our penguin dominated
$B\to\pi K$ modes. In the case of the former strategies, only small 
effects of new physics are expected since they are using essentially 
pure ``tree'' decays and no FCNC processes.   

\begin{figure}
\epsfysize=8.5cm
\centerline{\epsffile{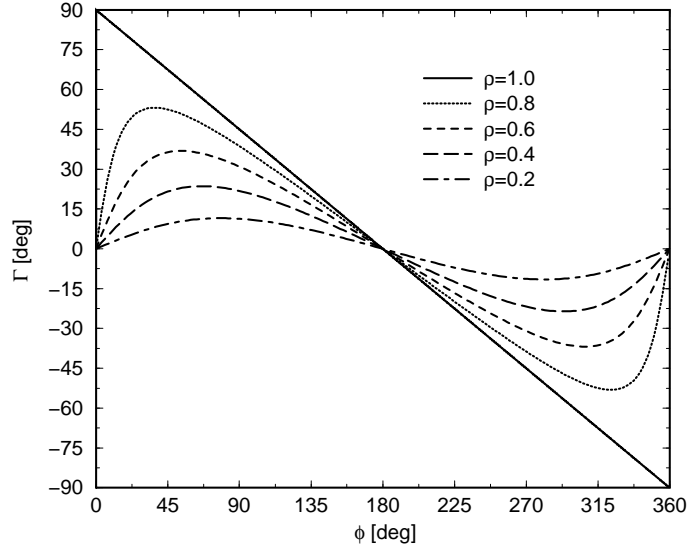}}
\caption[]{The dependence of $\Gamma$ on $\phi$ for various values of 
$\rho$.}
\label{fig:iso}
\end{figure}

In Fig.~\ref{fig:iso} we show the dependence of $\Gamma$ on $\phi$ for 
various values of $\rho$. Since (\ref{C-iso}) and (\ref{CPasym}) have 
exactly the same form as the corresponding SM expressions, the formalism 
developed in \cite{fm2} can be applied by making only the simple replacements 
$R\to R_{\rm exp}$, $r\to r_{\rm exp}$, $\gamma\to\gamma_{\rm exp}$. In 
particular, since $C_{\rm new}$ is still constrained between $-1$ and $+1$, 
the bounds on $r_{\rm exp}$ given in \cite{fm2} still remain valid. 
Consequently, whereas the overlap between (\ref{gamma-normal}) and 
(\ref{gamma-bound1}) can be increased, the possible problem related to 
the constraints on $r_{\rm exp}$ that we have pointed out in \cite{fm2}, 
namely that any reasonable estimate of this amplitude ratio is only 
marginally compatible with the present CLEO measurements, cannot be solved 
using this simple scenario. 

\boldmath
\subsection{New physics II:  $SU(2)$ isospin symmetry violation}
\unboldmath
A scenario of new physics which potentially cures the problem with the 
value of $r_{\rm exp}$ is one in which the new physics contributions couple 
differently to $d$- and $u$-quarks and hence violate isospin as is e.g.\ the 
case in models where a heavy boson is mediating additional $b\to s$ FCNC 
contributions. In order to implement this in a managable ``toy'' model, we 
assume  
\begin{equation}
\begin{array}{lcrcl}
\phi_d&=&\phi_u&=&\,0\\
\Delta_d&=&\Delta_u&=&\,0\,.
\end{array}
\end{equation}
The vanishing of these phases is certainly a restrictive assumption. It is 
only meant to demonstrate that (\ref{Def-cnew}) and (\ref{Def-rnew}) 
incorporate a possible solution to the potential consistency problem with 
$\gamma$ and $r_{\rm exp}$. In the case of our specific $SU(2)$-violating 
scenario of new physics, these expressions simplify considerably to
\begin{eqnarray}
C_{\rm new}&=&\left( \frac{1-\rho_u}{1-\rho_d} \right) 
              \cos \delta \, \cos \gamma\\
R_{\rm new}&=&\frac{\left(\rho_u-\rho_d\right)\left(2-\rho_u-\rho_d\right)}
{\left(1-\rho_d\right)^2}\,,
\end{eqnarray}
so that the bound $\gamma_0$ is given by
\begin{equation}
\gamma_0 =\arccos\left[\left(\frac{1-\rho_d}{1-\rho_u}\right)\,
C_{\rm new}\right].
\end{equation}  

\begin{figure}
\epsfysize=9cm
\centerline{\epsffile{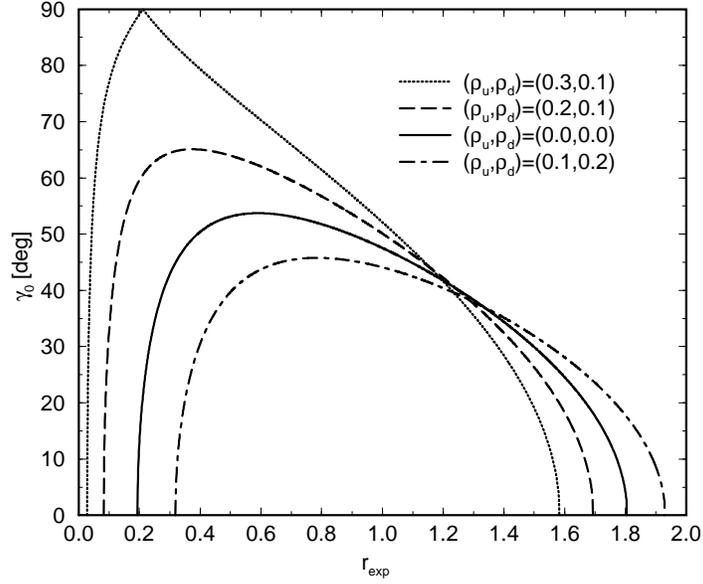}}
\caption[]{The dependence of $\gamma_0$ constraining the CKM angle $\gamma$
through (\ref{gamma-bound1}) on $r_{\rm exp}$ for a specific scenario of new 
physics discussed in the text and $R_{\rm exp}=0.65$ corresponding to the 
central values of the present CLEO measurements.}
\label{fig:gamma0-new}
\end{figure}

\begin{figure}
\epsfysize=9cm
\centerline{\epsffile{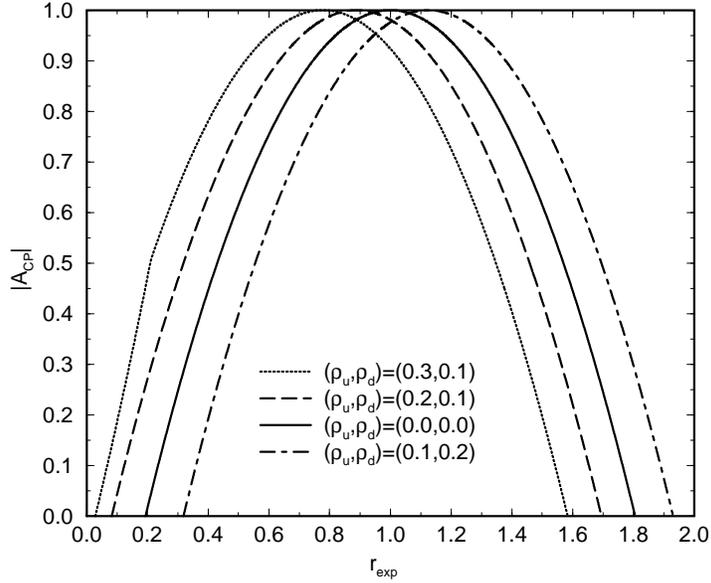}}
\caption[]{The dependence of the maximal value (\ref{CP-max}) of 
$|A_{\rm CP}^{\rm dir}(B^0_d\to\pi^-K^+)|$ on $r_{\rm exp}$ for 
$R_{\rm exp}=0.65$ and a specific scenario of new physics.}
\label{fig:ACP}
\end{figure}

Analogously to our recent paper \cite{fm2}, we show in 
Fig.~\ref{fig:gamma0-new} the dependence of $\gamma_0$ on $r_{\rm exp}$ for 
fixed $R_{\rm exp} = 0.65$  corresponding to the central values of the recent
CLEO measurements and various values of $(\rho_u,\rho_d)$. The new physics
affects also the direct CP-violating asymmetry of the decay $B_d \to 
\pi^\mp K^\pm$ as can be seen in Fig.~\ref{fig:ACP}, where we 
plot the maximal value  
\begin{equation}\label{CP-max}
\left|{\cal A}_{\rm CP}^{\rm dir}(B^0_d\to\pi^-K^+)\right|_{\rm max} = 
2\,\frac{r_{\rm exp}}{R_{\rm exp}}\left[\left(\frac{1-\rho_u}{1-\rho_d}\right)
-\left|C_{\rm new}\right|\right]
\end{equation}
of that asymmetry on $r_{\rm exp}$ for $R_{\rm exp}=0.65$ and various values 
of $(\rho_u,\rho_d)$. Looking at these figures, we observe that one can indeed 
incorporate smaller values of $r_{\rm exp}$ and larger values of 
$\gamma_0^{\rm max}$ in our simple isospin-breaking toy model if 
$\rho_u$ is larger than $\rho_d$. As far as the CP asymmetry is concerned, 
the curves for $\rho_u>\rho_d$ are shifted towards smaller 
$r_{\rm exp}$ compared to the SM case so that larger CP asymmetries can
be accommodated.

Also the SM has, however, an interesting source of isospin breaking, namely 
EW penguins. Since their contributions to the decays discussed in our paper
are expected to be negligibly small within the SM, the corresponding 
amplitudes have been neglected so far. In the next paragraph we will 
discuss their impact on our analysis in slightly more detail by assuming
a dramatic enhancement. 

\begin{figure}
\epsfysize=8.5cm
\centerline{\epsffile{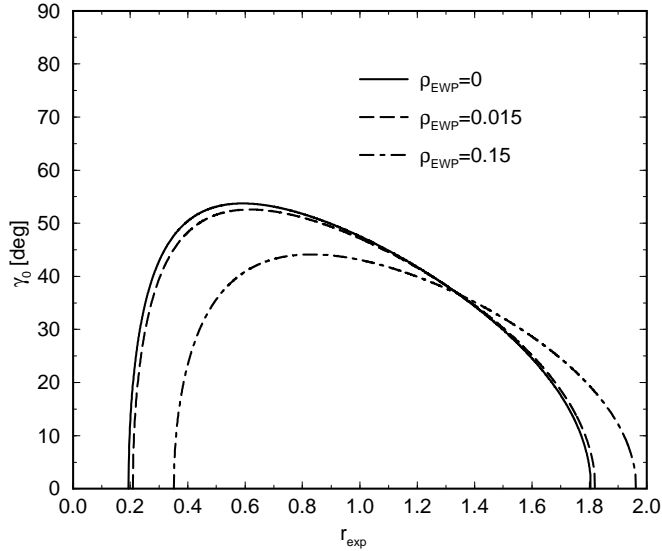}}
\caption[]{The dependence of $\gamma_0$ constraining the CKM angle $\gamma$
through (\ref{gamma-bound1}) on $r_{\rm exp}$ for $R_{\rm exp}=0.65$ in the
presence of enhanced EW penguins. The dashed line corresponds to the SM 
expectation for these operators.}
\label{fig:gamma0-EW}
\end{figure}

\boldmath
\subsection{New physics III: enhanced EW penguin contributions}\label{EWPs}
\unboldmath
The EW penguin contributions can easily be incorporated into our 
formulae by using 
\begin{eqnarray}
\rho_d=\frac{1}{3}\rho_{\rm EWP},&\mbox{}&\phi_d=0\nonumber\\
\rho_u=\frac{2}{3}\rho_{\rm EWP},&\mbox{}&\phi_u=\pi.
\end{eqnarray}
Moreover we assume $\Delta_d=\Delta_u=0$. The quantity $\rho_{\rm EWP}$ has 
been introduced already in (\ref{rho-EWP}) and measures the strength of the 
color-suppressed EW penguin contributions with respect to the QCD penguin 
amplitude. Based on the estimates in \cite{fm2} one finds $\rho_{\rm EWP}$ 
in the range of one to two percent. 

Let us assume that non-perturbative effects in the hadronic matrix 
elements of the corresponding EW penguin operators or
new physics effects give an enhancement of 
$\rho_{\rm EWP}$ by a factor of ${\cal O}(10)$ with respect to our simple 
estimates. This is shown in Fig.~\ref{fig:gamma0-EW}, where we have again 
used the central values of the CLEO measurements to fix $R_{\rm exp}$. From 
this figure we conclude that the SM contribution of the EW penguins is 
indeed negligible, and that a QCD enhancement tends to shift the bound  
$\gamma_0$ towards higher values of $r_{\rm exp}$. Hence it will probably 
not be 
able to cure the potential problem with the amplitude ratio $r_{\rm exp}$.
Moreover, an artificially enhanced EW penguin contribution will lower 
$\gamma_0$ leaving thus less overlap with the conventional bounds  
(\ref{gamma-normal}). For these considerations we have assumed that the
strong interaction effects enhancing the EW penguins will not drastically 
change the CP-conserving strong phases of the EW penguin amplitudes. This 
assumption is questionable and for a large phase shift, as e.g.\ 
$\Delta_d\approx\Delta_u \approx\pi$, the situation could as well reverse. 

\section{Conclusions}\label{concl}
It is generally accepted that penguin dominated decays of $B$ mesons
may be sensitive to new physics effects at a level which makes them 
interesting probes for non-SM effects. While this type of decays 
clearly will not be able to discriminate between different high energy 
scenarios, it still may turn out to be the first hint on physics 
beyond the SM. In \cite{fm2} we have pointed out that already combined 
branching ratios for $B^\pm\to\pi^\pm K$ and $B_d\to\pi^\mp K^\pm$ decays 
may lead to potential problems for the SM if the CLEO measurements should 
stabilize at their present central values. Firstly, the amplitude ratio 
$r_{\rm exp}$ of the current-current and the QCD penguin amplitudes is at 
the edge of compatibility with the SM, secondly the allowed range for 
$\gamma$ obtained from the combined $B\to\pi K$ branching ratios   
has only a small overlap with the conventional determination of this 
angle, and thirdly a future measurement of the CP asymmetry in 
$B_d\to\pi^\mp K^\pm$ may lead to a surprise. 

The most general ansatz introducing new physics into these decays turns 
out to have too many parameters to be useful. Hence some additional 
theoretical assumptions have to enter the game. A first restriction we have 
applied is to assume that the new physics contributions do not exhibit 
direct CP violation in the decays under consideration. We have given the 
expressions for the relevant observables in that particular case. They still 
involve six parameters related to new physics so that a general analysis 
becomes too clumsy. 

In order to proceed further, we have discussed three cases which we consider 
interesting, but which are of course quite restrictive. The first example 
assumes that the new physics contributions to the decays at hand are  
symmetric under $SU(2)$ isospin. In that case the experimentally determined 
value $\gamma_{\rm exp}$ of $\gamma$ is simply shifted by some angle 
$\Gamma$ depending on the new physics. Interestingly, while this scenario 
offers a solution of the potential problem with $\gamma$, it cannot solve the 
one related to $r_{\rm exp}$. The second example was taylored to tackle 
that problem. Here we assume that new physics breaks isospin. Making certain 
assumptions concerning strong and weak phases of the new physics, we have 
shown that appropriate isospin-breaking could indeed cure the potential 
problem with $r_{\rm exp}$ as well as the one with $\gamma$.

Finally we reconsider the isospin-breaking due to EW penguin operators 
which are already present in the SM. Estimates suffering from large hadronic 
uncertainties indicate that EW penguins should play a negligible role in 
the decays under consideration within the SM. An interesting observation 
is that the EW penguin contributions tend to shift the results such that the 
potential problems with consistency of the SM become worse. Even if for 
some reason the EW penguins become dramatically enhanced, 
they will not cure these potential problems unless rescattering effects 
yield large CP-conserving strong phase shifts as e.g.\ 
$\Delta_d\approx\Delta_u\approx\pi$. 

Although the hadronic uncertainties in $B\to\pi K$ modes, which are exclusive 
nonleptonic $B$ decays, are very large, these transitions may play an 
important role concerning the search for new physics. At first sight, this 
statement seems to be contradictory. However, the SM predicts the general 
phase structure of the corresponding decay amplitudes on solid ground. 
Moreover isospin symmetry of strong interactions -- working very well within 
the SM -- allows to derive relations among these decay amplitudes. 
Consequently, combining experimental data for $B\to\pi K$ in a clever way, 
the consistency of that description can be tested. Since experimental data on 
these decays is now starting to become available, certainly an exciting time 
is ahead of us and a future reduction of the presently large experimental 
uncertainties may shed light on physics beyond the SM.

\section*{Acknowledgments}

This work was supported by DFG under contract Ma 1187/7-1,2
and by the Graduierten\-kolleg ``Elementarteilchenphysik an Beschleunigern''.


\begin{thebibliography}{99}

\bibitem{ut}L.L. Chau and W.-Y. Keung, {\it Phys.\ Rev.\ Lett.}~{\bf 53} 
(1984) 1802; C. Jarlskog and R. Stora, {\it Phys.\ Lett.}~{\bf B208} 
(1988) 268.

\bibitem{ckm}N. Cabibbo, {\it Phys.\ Rev.\ Lett.}~{\bf 10} (1963) 531;
M. Kobayashi and K. Maskawa, {\it Prog.\ Theor.\ Phys.}~{\bf 49} (1972) 282.

\bibitem{rev}R. Fleischer, Univ.\ of Karlsruhe 
preprint TTP96-58, hep-ph/9612446, invited review article for publication 
in {\it Int.\ J.\ Mod.\ Phys.}~{\bf A}.

\bibitem{PAPIII}R. Fleischer, {\it Phys.\ Lett.}~{\bf B365} (1996) 399.

\bibitem{CLEO}J. Alexander, CLEO collaboration, talk given at the 2nd
International Conference on $B$ Physics and CP Violation, Honolulu, Hawaii,
24--27 March 1997; F. W{\"u}rthwein, CLEO collaboration, talk given at MPI 
Heidelberg and private communication; K. Ecklund, CLEO collaboration, talk 
given at Beyond the Standard Model V, Balholm, Norway, 29 April -- 4 May 1997.

\bibitem{fm2}R. Fleischer and T. Mannel, Univ.\ of Karlsruhe preprint 
TTP97-17, hep-ph/9704423.

\bibitem{bf-rev}A.J. Buras and R. Fleischer, TTP97-15, hep-ph/9704376,
to appear in {\it Heavy Flavours II}, Eds.\ A.J. Buras and M. Lindner 
(World Scientific, Singapore, 1997).

\bibitem{al}A. Ali and D. London, DESY 96-140, hep-ph/9607392.

\bibitem{NewPhysModels}For reviews see e.g.\ Y. Grossman, Y. Nir and
R. Rattazzi,  SLAC-PUB-7379, hep-ph/9701231; M. Gronau and D. London,
{\it Phys.\ Rev.}~{\bf D55} (1997) 2845; Y. Nir and H.R. Quinn,
{\it Ann.\ Rev.\ Nucl.\ Part.\ Sci.}~{\bf 42} (1992) 211.

\bibitem{appls}For applications see e.g.\ Y. Grossman and M.P. Worah, 
{\it Phys.\ Lett.}~{\bf B395} (1997) 241; M. Ciuchini et al., CERN-TH/97-47, 
hep-ph/9704274; R. Barbieri and A. Strumia, IFUP-TH 16/97, hep-ph/9704402. 

\bibitem{pens}J.-M. G\'erard and W.-S. Hou, {\it Phys.\ Lett.}~{\bf B253} 
(1991) 478; R. Fleischer, {\it Z. Phys.}~{\bf C58} (1993) 483; 
G. Kramer, W.F. Palmer and H. Simma, {\it Z. Phys.}~{\bf C66} 
(1995) 429. 

\bibitem{kagan}A. Kagan, {\it Phys.\ Rev.}~{\bf D51} (1995) 6196.

\bibitem{gw}M. Gronau and D. Wyler, {\it Phys.\ Lett.}~{\bf B265} (1991)
172; D. Atwood, I. Dunietz and A. Soni, {\it  Phys.\ Rev.\ Lett.}~{\bf 78}
(1997) 3257.

\bibitem{hawaii}For a recent review see e.g.\ R. Fleischer, TTP97-21, 
hep-ph/9705404, invited talk given at the 2nd International Conference on 
$B$ Physics and CP Violation, Honolulu, Hawaii, 24--27 March 1997, 
to appear in the proceedings.

\end{thebibliography}
\end{document}